\documentclass[11pt]{jhep}

\usepackage{amsmath}
\usepackage{amssymb}
\usepackage{graphicx}
\usepackage{axodraw}
\usepackage{mciteplus}

\usepackage[usenames]{color}
\let\normalcolor\relax      

\newcommand{\isotope}[2]{\phantom{}^{#2}\text{#1}}
\def\Hh{\isotope{H}{}}
\def\Hd{\isotope{D}{}}
\def\Ht{\isotope{H}{3}}
\def\He#1{\isotope{He}{#1}}
\def\Li#1{\isotope{Li}{#1}}
\newcommand{\stau}{\tilde{\tau}}
\newcommand{\mstau}{m_{\tilde{\tau}}}
\newcommand{\mgrav}{m_{3/2}}
\newcommand{\Mpl}{M_{\rm pl}}

\newcommand{\pfrac}[2]{\left( \frac{#1}{#2} \right)}

\newcommand{\unit}[1]{\; \mbox{#1}}

\newcommand{\fig}[1]{Fig.\;\ref{#1}}
\newcommand{\eq}[1]{Eq.\;(\ref{#1})}

\title{Gravitino dark matter and the lithium primordial abundance within a pre-BBN modified expansion}
\author{Sean Bailly\\
Laboratoire de Physique Th\'eorique LAPTH\\
Universit\'e de Savoie, CNRS (UMR 5108)\\
BP 110, F-74941 Annecy-Le-Vieux Cedex, France.\\
Email : \email{sean.bailly@lapp.in2p3.fr}}

\abstract{
\noindent

We present supersymmetric scenarios with gravitino LSP and stau NLSP in the case of a non-standard model of cosmology with the addition of a dark component in the pre-BBN era. In the context of the standard model of cosmology, gravitino LSP has drawn quite some attention as it is a good candidate for dark matter. It is produced in scattering processes during reheating after inflation and from the decay of the stau. With a long lifetime, the stau decays during Big Bang Nucleosynthesis. It is strongly constrained by the abundance of light elements but can however address the known "BBN lithium problem". It requires fairly massive staus $\mstau \gtrsim 1 \unit{TeV}$ and puts an upper bound on the reheating temperature $T_R \simeq 10^7\unit{GeV}$ which does not satisfy the requirements for thermal leptogenesis. For the non-standard cosmological scenario, the reheating temperature bound can be strongly relaxed $T_R\gg 10^9 \unit{GeV}$ and the lithium-7 problem solved with a stau typical mass of $\mstau \sim 600-700 \unit{GeV}$ and down to $\sim 400 \unit{GeV}$ with a very important dark component that could enable possible production and detection at the LHC.

\\
\\
}

\preprint{LAPTH-033/10}

\begin{document}

\section{Introduction}

From a cosmological perspective, supersymmetry is a promising theory beyond the Standard Model. Assuming the conservation of R-parity, the lightest supersymmetric particle (LSP), such as the neutralino or the gravitino, can be a candidate for dark matter. The gravitino acquires mass through the super-Higgs mechanism in broken local supersymmetry and which spans from few eV to TeV depending on the scenario. The gravitino LSP has been studied intensively \cite{Borgani:1996ag, Feng:2003xh, Feng:2003uy, Ellis:2003dn, Feng:2004zu, Feng:2004mt, Roszkowski:2004jd, Baltz:2001rq, Fujii:2002fv, Lemoine:2005hu, Jedamzik:2005ir, Cerdeno:2005eu, Buchmuller:2006nx, Steffen:2006hw, Kawasaki:2007xb, Pradler:2007is, Kersten:2007ab, Kawasaki:2008qe, Bailly:2008yy, Bailly:2009pe} as a candidate for dark matter.

The current dark matter density has been measured with great precision by the WMAP mission \cite{Komatsu:2008hk}, the five-year data giving at $3\sigma$ 
\begin{equation}
\Omega_{\rm DM} h^2 = 0.1099\pm 0.0124
\label{eq:WMAP}
\end{equation}
where $h$ is the reduced Hubble parameter. This result puts strong constraints on the production of gravitinos in the early Universe. Its relic density must lie below the upper limit to avoid an overclosure of the Universe. The lower bound on the dark matter is less stringent as one can have multi-component dark matter with gravitinos and other species.

There are two main gravitino production processes. On the one hand, the thermal production (TP) controlled by the post-inflation reheating temperature produces gravitinos in scattering processes during reheating \cite{Nanopoulos:1983up, *Ellis:1984eq, *Ellis:1984er, *Juszkiewicz:1985gg, Kawasaki:1994af, Moroi:1995fs, Bolz:1998ek, *Bolz:2000fu, Cyburt:2002uv, Kawasaki:2004yh, Kawasaki:2004qu, Roszkowski:2004jd, Kohri:2005wn, Cerdeno:2005eu, Jedamzik:2006xz, Pradler:2006qh, Rychkov:2007uq, Pradler:2007ne, Kawasaki:2008qe, Bailly:2009pe}. To avoid overclosure of the Universe, one obtains upper limits on the reheating temperature $T_R \lesssim 10^{6-8}\unit{GeV}$. The only experimental constraint on lower bounds for the reheating temperature comes from Big Bang Nucleosynthesis that can occur only if $T_{R} \geq 1 \unit{MeV}$. But the derived bound is somewhat below the required temperature for thermal leptogenesis $T_R \gtrsim 0.4-2\times 10^{10}\unit{GeV}$ \cite{Buchmuller:2004tu}. On the other hand, gravitinos are produced from the decay of supersymmetric particles. Since gravitinos only interact gravitationally with the other particles, these interactions are suppressed by the Planck mass and are therefore very small (in the limit of our considered gravitino masses, for very light gravitinos, the goldstino component increases the interactions). While the Universe cools down, SUSY particles would decay preferably to the next to lightest supersymmetric particle (NLSP) which would then decay to the gravitino. This process is called the non-thermal production (NTP) of gravitinos \cite{Borgani:1996ag, Feng:2003xh, Feng:2003uy, Ellis:2003dn, Feng:2004zu, Feng:2004mt}. The contribution of both NTP and TP processes contributing to the total gravitino relic density was performed in \cite{Roszkowski:2004jd, Cerdeno:2005eu, Bailly:2008yy, Bailly:2009pe} with the addition of Big Bang Nucleosynthesis (BBN) constraints to solve the lithium problems. In standard cosmology, the Big Bang Nucleosynthesis (BBN) is a very predictive model yielding the abundance of light elements produced in the early Universe. These predictions are confronted to observations and are in great compatibility for deuterium, in quite good agreement for helium-4 but present strong discrepancies for lithium-7 and lithium-6. Solutions to the lithium problems are investigated on an astrophysics perspective or on the particle physics side by assuming the decay of relic particles during BBN. Solutions for the lithium problems were studied in specific models such as CMSSM or GMSB models with stau or neutralino NLSP decaying to the gravitino LSP \cite{Jedamzik:2004er, Jedamzik:2005dh, Steffen:2006hw, Pradler:2007is, Bailly:2008yy, Bailly:2009pe}. The NLSP is long-lived due to the Planck mass suppressed coupling to the gravitino and can decay during the BBN at times above one second. In the CMSSM framework with gravitino LSP and stau NLSP, it was shown that it is possible to solve both lithium problems and obtain the dark matter relic density from gravitino thermal and non-thermal production \cite{Jedamzik:2005dh, Bailly:2008yy, Bailly:2009pe}. The downsides are the necessity for quite small reheating temperature and a fairly heavy spectrum.

In this study we consider CMSSM scenarios with a stau NLSP, a gravitino LSP produced from TP and NTP and we assume a non-standard cosmological history. We introduce a modified expansion rate due to the energy density of a dark component in the pre-BBN era following the model presented by Arbey and Mahmoudi \cite{Arbey:2008kv, Arbey:2009gt}. The consequence of such a modification of the Hubble parameter is a less efficient production of gravitino in the thermal process and an early freeze-out of the NLSP leading to higher abundance of the NLSP. The constraints on the reheating temperature are therefore less stringent and allow higher values compatible with thermal leptogenesis. The higher abundance of NLSP allows to solve the lithium problems with lighter masses for the NLSP than in the standard scenario. Consequently these supersymmetric spectra could be accessible to collider experiments.

This paper is organised as follows, in section \ref{sec:model} we describe the model with a pre-BBN expansion rate. In sections \ref{sec:TP} and \ref{sec:NTP} we present the calculation of the gravitino relic density through thermal and non-thermal production followed in section \ref{sec:BBN} by the resolution of the lithium problems from the decay of long-lived stau.

\section{\label{sec:model}Pre-BBN modified expansion}

Observations of the early Universe are mostly reliable for temperatures up to $T \sim 1 \unit{MeV}$ which correspond to the beginning of the Big Bang Nucleosynthesis. From that moment the Universe cooled down and its evolution is described by the standard model of cosmology, the hot Big Bang model. It is constrained by many observations such as the abundance of light elements, the cosmic microwave background radiation or the formation of large scale structures. The Universe must be dominated by radiation between BBN and the recombination followed by an era dominated by matter. On the contrary the evolution of the Universe in the pre-BBN era is unclear. There are no strong constraints on its composition. This open question can be related to the nature of dark energy, responsible today for the accelerated expansion of the Universe which is still unknown and could play an important role in the pre-BBN era. This is indeed the case in dark fluid models \cite{Arbey:2006it} or quintessence models \cite{Ratra:1987rm}. These dark energy components contribute to the Hubble parameter leading to faster or slower expansion rates (negative effective energies can be obtained in extra-dimension models \cite{Barger:2003zg, *Okada:2004nc}).

To model the effects of a modified expansion in the pre-BBN era, we will use the parametrisation given by Arbey and Mahmoudi \cite{Arbey:2008kv} adding a new dark density to the radiation density depending on temperature and characterising a fluid in adiabatic expansion
\begin{equation}
\rho_D(T)=\rho_D(T_{\rm BBN})\pfrac{T}{T_{\rm BBN}}^{n_D}
\end{equation}
where $n_D$ is a constant describing the density behaviour and the reference temperature $T_{\rm BBN}=10\unit{MeV}$ is chosen higher than the beginning of Big Bang Nucleosynthesis and neutrino freeze-out in order to assure a radiation dominated era for BBN. The equation of state is $w_D=P_D/\rho_D$ where $P_D$ is the pressure of the fluid, $n_D$ and $w_D$ are related by $n_D=3(w_D+1)$ (more details can be found in \cite{Ratra:1987rm}). The value $n_D=3$ ($w_D=0$) corresponds to a matter density, $n_D=4$ ($w_D=1/3$) to a radiation density and $n_D=6$ ($w_D=1$) corresponds to quintessential kination (quintessence with kinetic energy dominating potential energy \cite{Salati:2002md}). Higher values of $n_D$ can be reached in models with a decaying scalar field.

As we require a radiation dominated era for BBN, we introduce the parameter
\begin{equation}
\kappa_D=\frac{\rho_D(T_{\rm BBN})}{\rho_{\rm rad}(T_{\rm BBN})} \ll 1
\end{equation}
where $\rho_{\rm rad}$ is the radiation density
\begin{equation}
\rho_{\rm rad}(T) = g_{\rm eff}(T)\frac{\pi^2}{30}T^4
\end{equation}
$g_{\rm eff}$ is the effective number of radiation degrees of freedom. Higher values of $\kappa_D$ increases the length of the dark component domination era. For $\kappa_D \rightarrow 1$, radiation and dark component are co-dominant at $T_{\rm BBN}$ leading to perturbations in the production of light elements.

For the two parameters controlling this toy model of dark component, we will restrict our study to values $0<\kappa_D \ll 1$ and $4\leq n_D \leq 8$ as in \cite{Arbey:2008kv}. 

The Friedmann equation reads
\begin{equation}
H^2 = \frac{8\pi G}{3}\left( \rho_B + \rho_D\right)
\label{eq:Hubbleparam}
\end{equation}
where the energy of the background $\rho_B=\rho_{\rm rad} + \rho_{m}$ was dominated by radiation in the past and is now dominated by matter. For our purpose $\rho_B\simeq \rho_{\rm rad}$. Furthermore once BBN starts, the dark component becomes negligible and will have little effect on the expansion rate from that time on. The addition of the dark component in \eq{eq:Hubbleparam} will have important consequences for the calculation of the relic density of dark matter.

\section{\label{sec:TP}Thermal production of gravitino}

During the reheating era after inflation, scattering processes producing one gravitino can become very efficient. Bolz et al. \cite{Bolz:1998ek, *Bolz:2000fu} have considered a consistent thermal field theory approach for supersymmetric strong interactions. Pradler and Steffen have taken into account the full Standard Model gauge group $SU(3)_c\times SU(2)_L\times U(1)_Y$ using the same procedure \cite{Pradler:2006qh, Pradler:2007ne}. Another approach \cite{Rychkov:2007uq} leads up to a factor 2 difference with the previous study. Although, this approach is more complete, we will follow the procedure used by Pradler and Steffen for implementation purposes.

The Boltzmann equation describing the evolution of gravitino abundance reads
\begin{equation}
\frac{dn_{3/2}}{dt}+3Hn_{3/2} = C_{3/2}
\end{equation}
with $n_{3/2}$ the gravitino number density and $C_{3/2}$ the collision term describing the production and destruction of gravitinos. The Hubble parameter \eq{eq:Hubbleparam} can be rewritten
\begin{equation}
H(T)=\sqrt{\frac{\pi^2}{90}}\frac{g_{\rm eff}^{1/2}T^2}{\Mpl}\left( 1+ \kappa_D \pfrac{T}{T_{\rm BBN}}^{n_D-4}\right)^{1/2}
\end{equation}
where $\Mpl=2.4\times 10^{18}\unit{GeV}$ is the reduced Planck mass. Using the definition $Y_{3/2}=n_{3/2}/s$ where $s$ is the entropy and assuming that previous to reheating all gravitino density is diluted by inflation, we have $Y_{3/2}(T_R)=0$ with $T_R$ the reheating temperature. Integrating the Boltzmann equation between $T_R$ and $T_{\rm BBN}$, and assuming that all thermal production of gravitino is negligible at BBN times leads to
\begin{equation}
Y_{3/2}(T_{\rm BBN})=-\int_{T_R}^{T_{\rm BBN}} dT \frac{C_{3/2}(T)}{s(T)H(T)T}.
\label{eq:Y32general}
\end{equation}
The entropy density is
\begin{equation}
s= \frac{2 \pi^2}{45}h_{\rm eff}T^3.
\end{equation}
Thermal equilibrium is assumed for all supersymmetric particles during reheating therefore $g_{\rm eff}=h_{\rm eff}=915/4$.

Pradler and Steffen \cite{Pradler:2006qh} have calculated the collision terms including scattering processes producing one gravitino in strong and electroweak interactions
\begin{equation}
C_{3/2}(T)=\sum_{\alpha=1}^3 \left(1+\frac{M_{\alpha}^2}{3m_{3/2}^2} \right)\frac{3\zeta(3)T^6}{16\pi^3\Mpl^2}c_{\alpha} g_{\alpha}^2 \ln \pfrac{k_\alpha}{g_\alpha}.
\end{equation}
The sum is over the gauge groups with the calculated coefficients $c_\alpha=\{11, 27, 72\}$ and $k_\alpha=\{1.266, 1.313, 1.271\}$. The gaugino masses and the gauge couplings are respectively $M_{\alpha}$ and $g_\alpha$.

For $\kappa_D=0$, we recover the standard expression for the Hubble parameter. One obtains the result obtained by Pradler and Steffen by solving \eq{eq:Y32general} with $T_{\rm BBN} \ll T_R$
\begin{equation}
Y_{\rm PS}(T_{\rm BBN})=\sum_{\alpha=1}^3 \left(1+\frac{M_{\alpha}^2}{3\mgrav^2}\right) y_{\alpha} g_{\alpha}^2 \ln\pfrac{k_\alpha}{g_\alpha}\pfrac{T_R}{10^{10} \unit{GeV}}
\label{eq:Ypradler}
\end{equation}
with $y_\alpha=\{0.65, 1.6, 4.3\}\times 10^{-12}$. One can note the dependence on the reheating temperature, the gravitino mass and on the gaugino masses. Higher temperature and gaugino masses or smaller gravitino masses lead to a larger production of gravitinos.

For the case $\kappa_D \neq 0$, we take into account the effects of the dark component on the Hubble parameter. We get
\begin{equation}
Y_{3/2}(T_{\rm BBN})=2.08\times 10^{-23} \sum_{\alpha=1}^3 \left(1+\frac{M_{\alpha}^2}{3\mgrav^2}\right) c_\alpha g_{\alpha}^2 \ln\frac{k_\alpha}{g_\alpha}\int_{T\rm BBN}^{T_R} \frac{dT}{\left[ 1+\kappa_D\pfrac{T}{T_{\rm BBN}}^{n_D-4}\right]^{1/2}}
\label{eq:YTP}
\end{equation}
If $n_D=4$, the integral is independent of $T$ and the yield of gravitino is
\begin{equation}
Y_{3/2}(T_{\rm BBN})=\frac{1}{1+\kappa_D}Y_{\rm PS}(T_{\rm BBN})
\label{eq:nD4}
\end{equation}
The result is simply proportional to $Y_{\rm PS}(T_{\rm BBN})$ and for $\kappa_D\rightarrow 1$, we get $Y_{3/2}(T_{\rm BBN}) \rightarrow Y_{\rm PS}(T_{\rm BBN})/2$, and for $\kappa_D=0$ we recover the result of the standard scenario. The contribution of the dark component to the Hubble parameter tends to weaken the efficiency of gravitino production.

When $n_D > 4$, integration of \eq{eq:YTP} can again be expressed with \eq{eq:Ypradler} multiplied by an hypergeometric function
\begin{equation}
Y_{3/2}(T_{\rm BBN})=Y_{\rm PS}(T_{\rm BBN})\times{}_2F_1\left(1/N,1/2;1+1/N;-\kappa_D \pfrac{T_{R}}{T_{\rm BBN}}^N\right)
\end{equation}
with $N=n_D-4$. Since $\kappa_D>0$, the hypergeometric function ${}_2F_1$ takes values in the range $]0,1[$ so the gravitino production is always smaller than the standard production. It can also be directly seen from \eq{eq:YTP}. The relic density of thermally produced gravitino is
\begin{equation}
\Omega_{3/2}^{\rm TP}h^2 = 2.742 \times 10^8 \pfrac{\mgrav}{1\unit{GeV}}Y_{3/2}(T_{\rm BBN})
\label{eq:RelicTP}
\end{equation}

In all cases, we find that the gravitino production is proportional to the standard result calculated by Pradler and Steffen multiplied by a coefficient between 0 and 1. The addition of the dark component will induce the reduction of the gravitino thermal production. The production is smaller as the dark component dominates, that is if one takes large values of $\kappa_D$ or $n_D$. This can be easily understood: the Hubble parameter is much larger than in the standard scenario meaning that one exits the reheating period much faster leaving less time to produce gravitinos.

\begin{figure}
\centering \includegraphics[width=10cm]{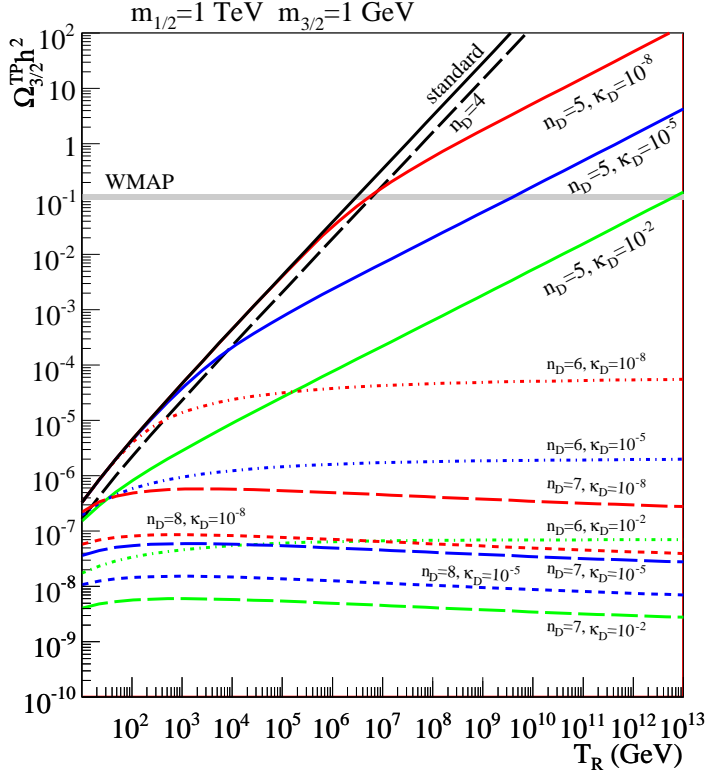}
\caption{\label{fig:thermal}Thermal production for different gravitino masses and \{$\kappa_D$,$n_D$\} at $m_{1/2}=1\unit{TeV}$ and $m_{3/2}=1\unit{GeV}$. The black line is the standard case corresponding to the maximal production of gravitino, the black dashed line has $n_D=4$. Then full lines have $n_D=5$, short-long dashed line $n_D=6$, long dashed lines $n_D=7$ and short dashed lines $n_D=8$, red, blue and green correspond respectively to $\kappa_D=\{10^{-8},10^{-5},10^{-2}\}$ }
\end{figure}

The thermal production of gravitino \fig{fig:thermal} shows the gravitino relic density with respect to the reheating temperature at $m_{1/2}=1\unit{TeV}$ and a gravitino mass $\mgrav=1\unit{GeV}$. We recall from \eq{eq:Ypradler} that the thermal production is more efficient for lighter gravitino masses. The grey area corresponds to the WMAP value of dark matter relic density and puts constraints on the reheating temperature. To avoid an overproduction of dark matter in the standard cosmological scenario (black continuous line),  the reheating temperature must lie below $T_{R}\sim 10^{6-7}\unit{GeV}$. The constraint becomes more severe for lighter gravitino masses but can go up to $T_{R}\sim 10^9\unit{GeV}$ for a gravitino mass of $m_{3/2}\simeq 100\unit{GeV}$ and $m_{1/2}=1\unit{TeV}$. Higher values of $m_{1/2}$ also increase the production of gravitinos. These limits on the reheating temperature are quite low for leptogenesis requirements ($T_R\sim 2\times 10^9 \unit{GeV}$). One must either turn to other mechanisms for baryogenesis (for instance \cite{Davidson:2000dw}) or limit the production of gravitino. This can be achieved in our model by inducing a modification of the expansion rate. Models with different values for $\kappa_D$ and $n_D$ are presented \fig{fig:thermal}. As a general conclusion, they all produce less gravitinos than the standard scenario, as expected, and therefore weaken the constraint on the reheating temperature. The black dashed line corresponds to models with $n_D=4$ as in \eq{eq:nD4} in the limit case of $\kappa_D\rightarrow 1$. The gravitino relic density is proportional to the Pradler-Steffen solution with a factor between 0.5 and 1, so the constraints on the reheating temperature are still close to the standard case. The gravitino production will be reduced with respect to the amount of dark component. For $n_D=5$ and $\kappa_D=10^{-5}$ the reheating temperature limit is $\sim 10^{9} \unit{GeV}$ and for higher values of $n_D$ the thermal contribution to the relic density is negligible, yielding no limit on the reheating temperature. It is now possible to satisfy the lower bound on the reheating temperature deriving from thermal leptogenesis.

Gravitino can also be produced from other mechanisms that must be added to the suppressed thermal production in order to obtain the dark matter relic density.

\section{\label{sec:NTP}Non-thermal production of gravitino}

This process comes from the decay of supersymmetric particles into the gravitino produced in the early Universe before the beginning of BBN. Due to the Planck mass suppression in the gravitino couplings, SUSY particles decay to the NLSP which then decays to the gravitino. One can assume a standard calculation of the abundance of the NLSP as if it was stable by solving the Boltzmann equation \cite{Gondolo:1990dk}. Then one can assume that each NLSP will produce one gravitino and the gravitino relic density is related to the abundance of NLSP scaled by the mass ratio
\begin{equation}
\Omega_{3/2}^{\rm NTP}h^2 = \frac{m_{3/2}}{m_{\rm NLSP}} \Omega_{\rm NLSP}h^2
\label{eq:OmegaNTP}
\end{equation}
The unknown abundance of NLSP needs to be calculated. Its computation is based on the solution of the Boltzmann equation
\begin{equation}
\frac{dn}{dt}=-3Hn -\left< \sigma_{\rm eff} v\right>(n^2-n_{\rm eq}^2)
\label{eq:Boltzmann}
\end{equation}
where $n$ is the number density of all supersymmetric particles, $n_{\rm eq}$ is their equilibrium density, and $\left< \sigma_{\rm eff} v\right>$ is the thermal average of the annihilation rate of the supersymmetric particles to the Standard Model particles. The numerical computation has been automated under the standard scenario in programs such as MicrOMEGAs \cite{Belanger:2006is} or DarkSUSY \cite{Gondolo:2004sc} with absence of entropy production and decay of relic particles.

The Boltzmann equation \eq{eq:Boltzmann} shows that the modification of the Hubble parameter with the introduction of the dark component will have strong impact on the freeze-out of the NLSP and its abundance. This has been studied in the case of the neutralino LSP \cite{Salati:2002md}. Since the Hubble parameter is larger with the dark component, the freeze-out occurs earlier leaving a larger relic density because less annihilations were performed. Similar conclusions are drawn in the stau NLSP case.

We use the usual ratio $x=m_{\rm NLSP}/T$ of the NLSP mass over temperature and combining \eq{eq:Hubbleparam} and \eq{eq:Boltzmann} it yields
\begin{equation}
\frac{dY}{dx}=-\sqrt{\frac{\pi}{45}}\frac{g_{*}^{1/2}m_{\rm NLSP}}{x^2}\left( 1+ \frac{\rho_D(T)}{\rho_{\rm rad}\frac{\pi^2}{30}T^4}\right)^{-1/2} \left< \sigma_{\rm eff}v\right> \left(Y^2-Y_{\rm eq}^2\right)
\label{eq:BoltzY}
\end{equation}
with
\begin{equation}
g_{*}^{1/2}=\frac{h_{\rm eff}}{\sqrt{g_{\rm eff}}}\left(1+\frac{T}{3h_{eff}}\frac{dh_{\rm eff}}{dT}\right)
\end{equation}
The standard result is obtained with $\rho_D \rightarrow 0$.

The abundance of the NLSP is then obtained by integrating \eq{eq:BoltzY} between $x=0$ and $x=m_{\rm NLSP}/T_0$ where $T_0=2.726 \unit{K}$ is the temperature of the Universe today. The yield is related to the relic density as in \eq{eq:RelicTP}
\begin{equation}
\Omega_{\rm NLSP} h^2 = 2.742 \times 10^8 \pfrac{m_{\rm NLSP}}{1\unit{GeV}}Y(T_0)
\label{eq:OmegaNLSP}
\end{equation}

Combining \eq{eq:OmegaNTP} and \eq{eq:OmegaNLSP}, the non-thermal part of the gravitino relic density is proportional to the gravitino mass and to $Y(T_0)$. The latter is numerically computed using a modified version of MicrOMEGAs 2.2 \cite{Belanger:2006is} with SuSpect 2.41 \cite{Djouadi:2002ze}.

\begin{figure}
\begin{center}
\centering
\includegraphics[width=7cm]{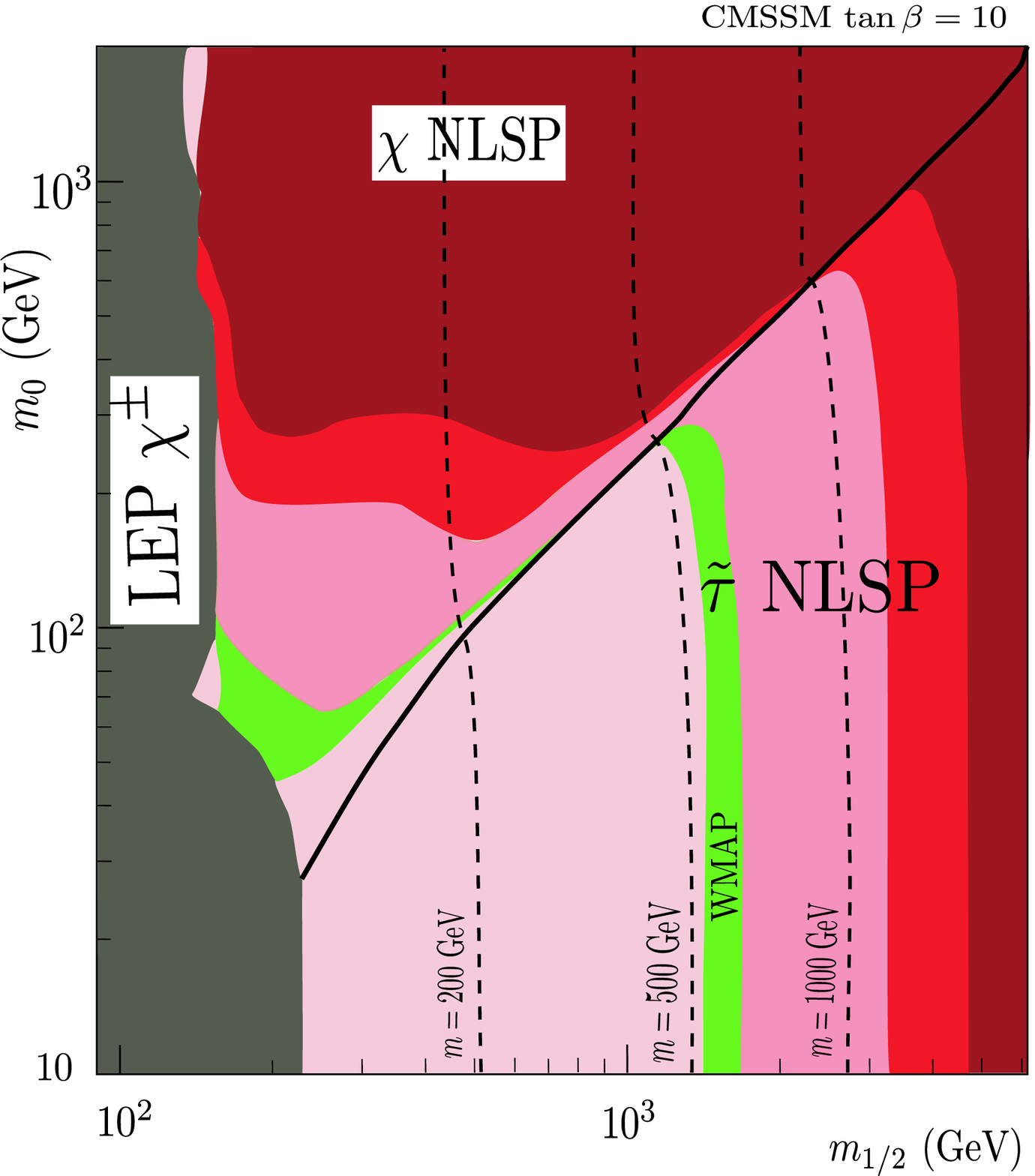}
\qquad
\includegraphics[width=7cm]{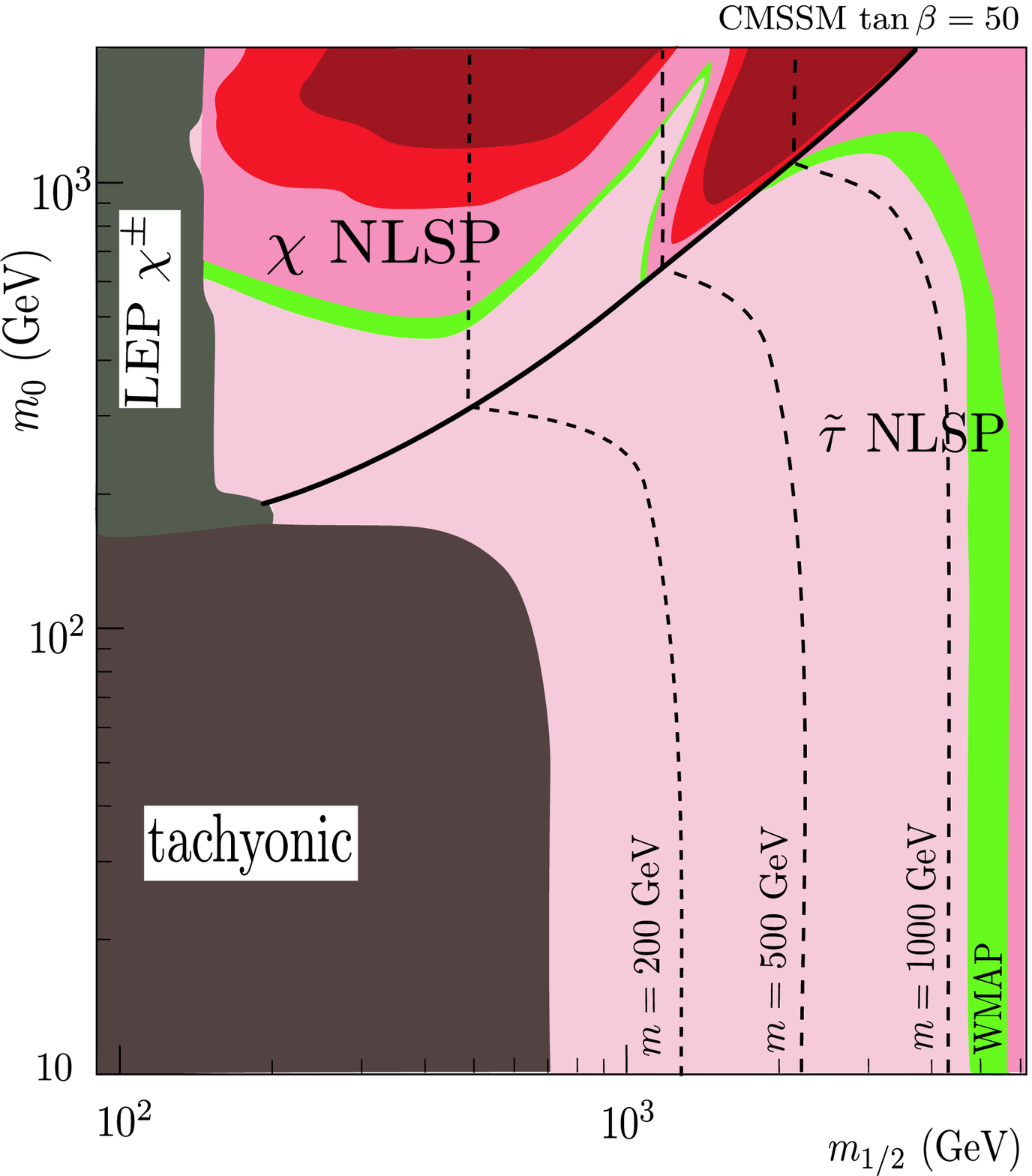}
\caption{\label{Fig:CMSSM}Contours of $\Omega_{\rm NLSP} h^2$ in the $m_0-m_{1/2}$ plane of the CMSSM for $\tan\beta=10$ (left) and $\tan\beta=50$ (right). In green the region where  $\Omega_{\rm NLSP} h^2$ is compatible with WMAP measurements. In light pink the region where $\Omega_{\rm NLSP}h^2 < 0.091$, in pink $0.128 < \Omega_{\rm NLSP}h^2 < 0.5$, in red $0.5 < \Omega_{\rm NLSP}h^2 < 1$ and in dark red $\Omega_{\rm NLSP}h^2>1$. The grey area noted ``LEP" is an exclusion contour coming from lower bounds on SUSY masses from LEP and ``tachyonic" signals that there are no RGE solutions. The dashed lines are contours of the NLSP masses.}
\end{center}
\end{figure}

In the context of the CMSSM, the NLSP is mainly the lighter stau or the lightest neutralino. In this paper we will only focus on the stau case. First we will recall different results concerning the relic density of stau in the standard scenario without dark component. CMSSM depends on a small number of parameters, namely $m_{1/2}$ the gaugino mass at the GUT scale, $m_0$ the scalar mass at the GUT scale, $A_0$ the trilinear coupling at the GUT scale that we take $A_0=0$, $\tan \beta$ ratio of the Higgs vacuum expectation values and the sign of $\mu$ the Higgs mass parameter. We scan the parameter space with $m_{1/2}\in [100,6000]\unit{GeV}$, $m_0\in[10,2000]\unit{GeV}$, $A_0=0$, $\mu>0$ and $\tan \beta=10$ or $50$. \fig{Fig:CMSSM} presents masses and abundances of the stau (and neutralino) in this model assuming a standard cosmological history. The green area noted WMAP indicates a relic density compatible with \eq{eq:WMAP}. It is mostly relevant to the neutralino as it could be a candidate for dark matter if the gravitino is heavier. For the stau NLSP \fig{Fig:CMSSM} emphasizes with respect to \eq{eq:OmegaNTP} that for lower values of $m_{1/2}$, NTP alone cannot produce dark matter and one must consider TP contribution. We have also added exclusion contours from LEP coming mainly from the lower bound on the chargino mass and at $\tan\beta=50$ a region where renormalization group equations do not give correct solutions for the SUSY masses. We have also indicated limits on the NLSP mass. Roughly, at $m_{1/2}\gg m_0$ the stau is the NLSP. Its mass increases with the gaugino mass. The stau mass can be analytically approximated by \cite{deBoer:1994he} at weak scale
\begin{equation}
\mstau^2 = m_{0}^2 + 0.15 m_{1/2}^2 - 0.23 \cos(2\beta)M_{Z}^2
\end{equation}
The relic density the stau would have if it had not decayed to the gravitino reads \cite{Asaka:2000zh}
\begin{equation}
\Omega_{\tilde{\tau}}h^2 = (2.2-4.4)\times 10^{-1} \pfrac{\mstau}{1\unit{TeV}}^2
\label{eq:estimrelicstau}
\end{equation}
The coefficient in \eq{eq:estimrelicstau} depends on the importance of slepton coannihilation. In the region $m_{1/2} \gg m_0$, it is clear that the stau relic density increases with $m_{1/2}$ independently of $m_0$. Dependence on $m_0$ becomes important in the stau-neutralino coannihilation region. For a given value of $m_{1/2}$, the abundance of stau increases due to a smaller cross-section from coannihilation processes.
At $\tan\beta=50$, the parameter region with stau NLSP is wider and the relic density is smaller due a larger left-right mixing and enhanced stau-Higgs coupling.

\begin{figure}
\centering
\includegraphics[width=7.2cm]{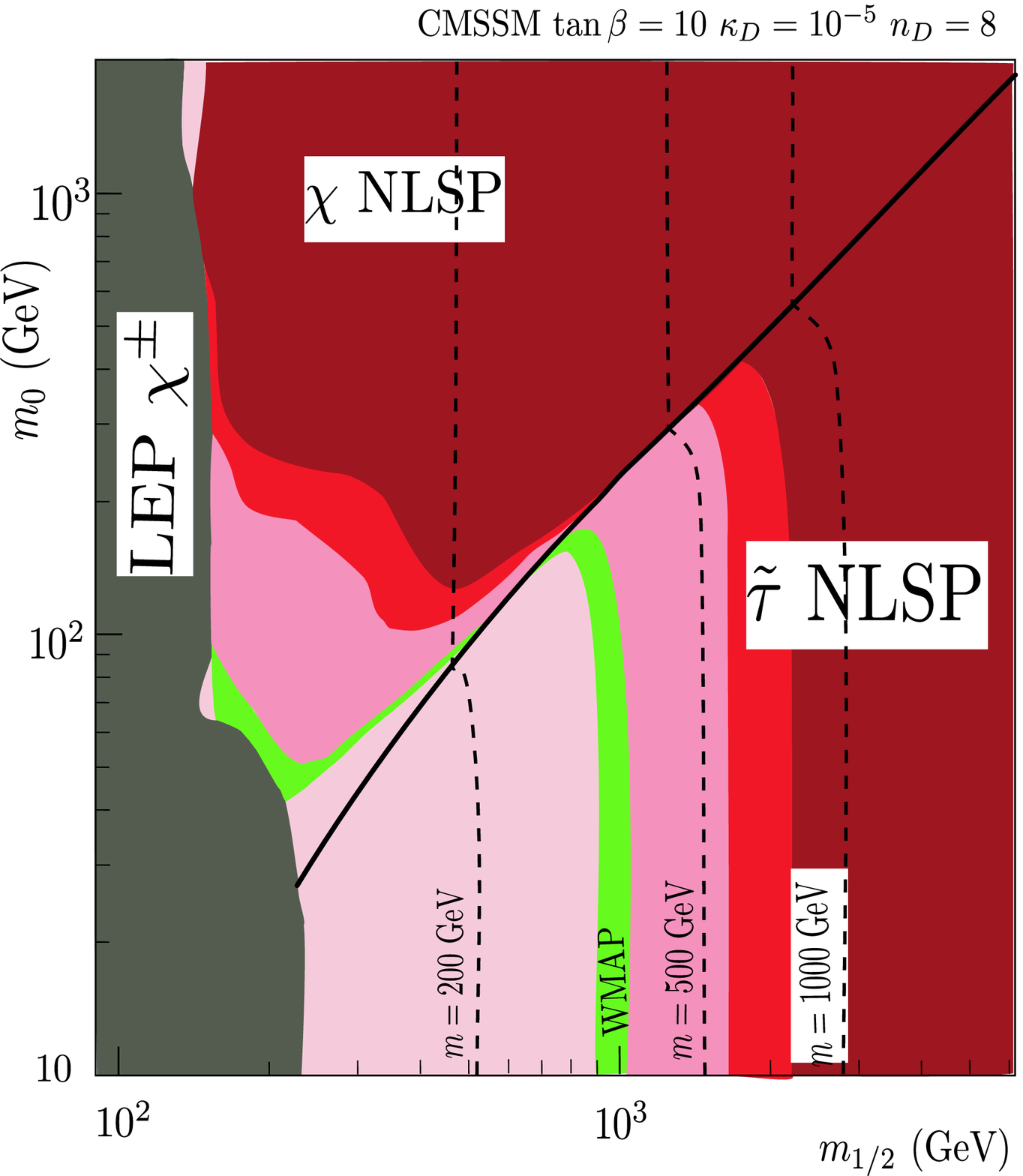}
\quad
\includegraphics[width=7.2cm]{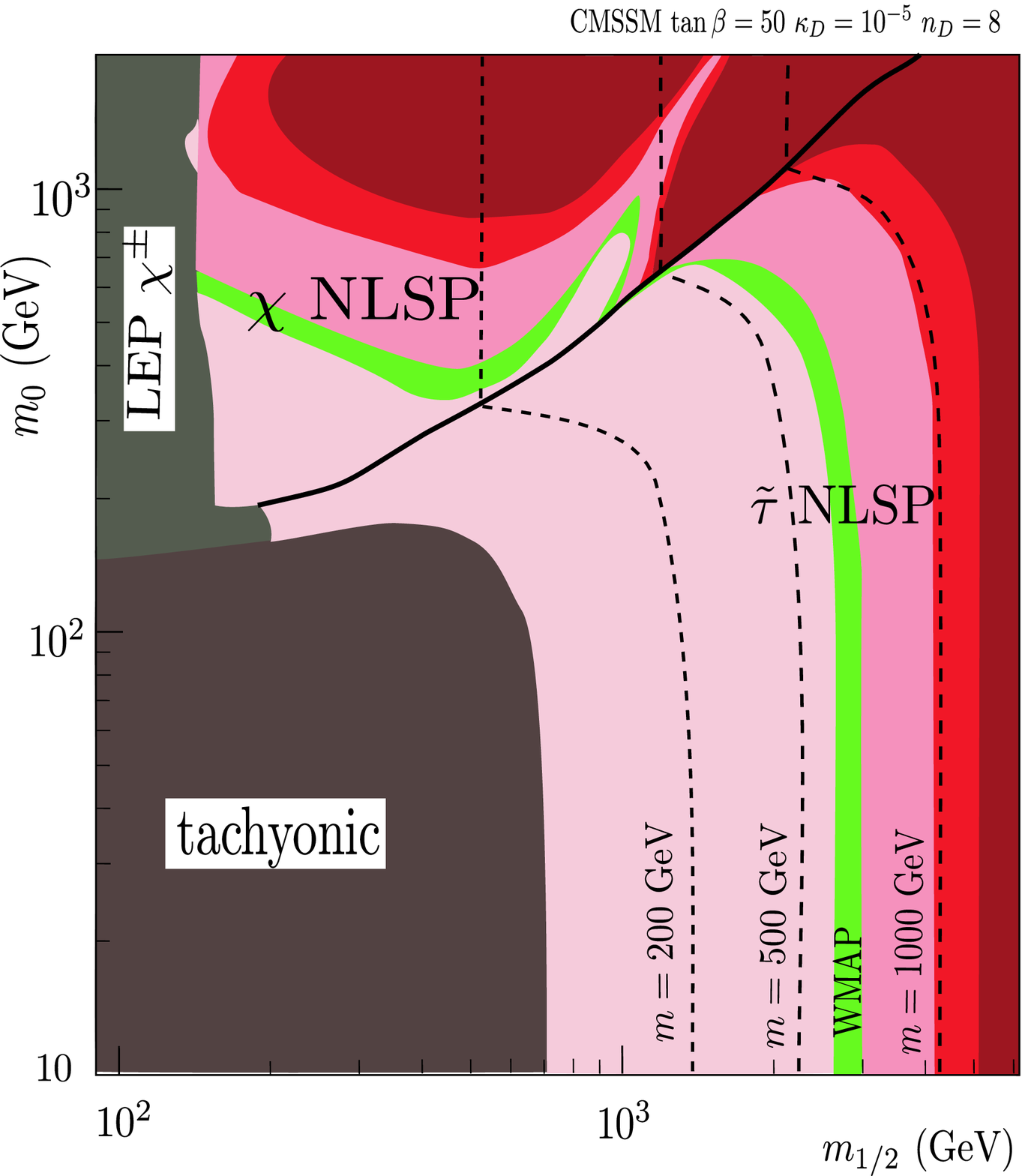}
\caption{\label{Fig:CMSSM_modgrav}Relic density of NLSP stau and neutralino for $\kappa_D=10^{-5}$, $n_D=8$ and $\tan\beta=10$ (left) or $\tan\beta=50$ (right), same colours as in \fig{Fig:CMSSM}}
\end{figure}

\begin{figure}
\centering
\includegraphics[width=10cm]{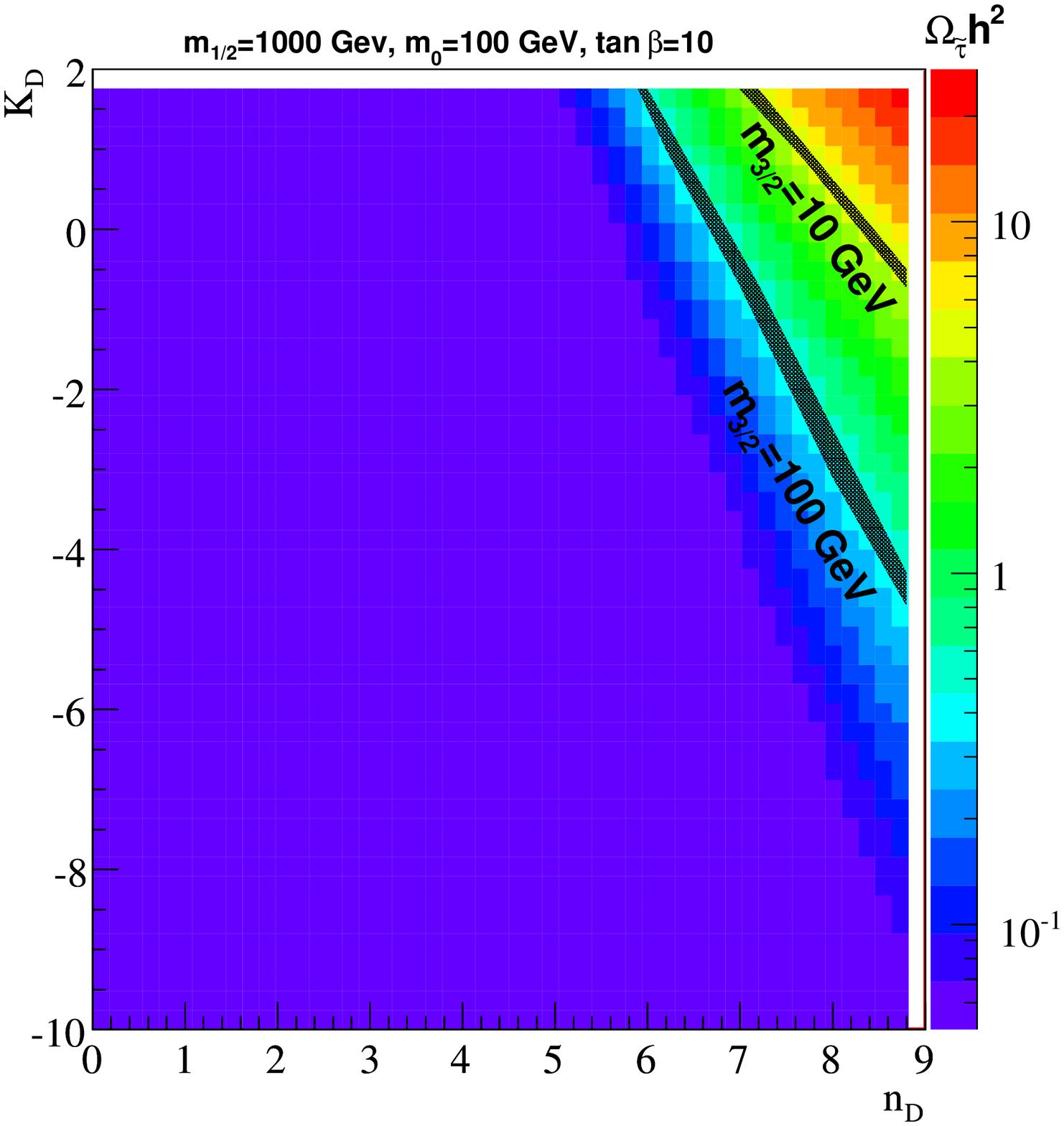}
\caption{\label{Fig:relic_modgrav}Relic density of a stau NLSP in the $\kappa_D-n_D$ plane for $\tan\beta=10$, $m_{1/2}=1000$ GeV and $m_0=100$ GeV corresponding to a stau mass $\mstau\sim 370 \unit{GeV}$. Lines indicate gravitino relic density from NTP compatible with WMAP bounds on dark matter relic density for gravitino masses $m_{3/2}=10, 100 \unit{GeV}$}
\end{figure}

Adding the dark component, we increase the value of the Hubble parameter with a positive effective energy density. Therefore we expect the freeze-out to occur earlier since the condition to freeze-out is approximately $H\gtrsim \Gamma$ with the reaction rate $\Gamma \propto n \left<\sigma v\right>$. If the freeze-out occurs earlier, less annihilation occurs and the relic density must be higher.

As in the standard scenario, \fig{Fig:CMSSM_modgrav} reproduces the relic density of the  NLSP for $\tan\beta=10$ (left) and $\tan\beta=50$ (right) and the dark component parameters are set to $\kappa_D=10^{-5}$ and $n_D=8$.

Comparing with \fig{Fig:CMSSM}, one can immediately assess the increase in relic density for the NLSP. For example, a relic density of $\Omega_{\stau}h^2\sim 0.1$ was achieved for $m_{1/2}\simeq1500\unit{GeV}$ at $\tan\beta=10$ in the standard cosmological scenario and at $m_{1/2}\simeq 900-1000\unit{GeV}$ in this modified cosmological scenario. The same observations can be made at higher values of $\tan\beta$. The direct consequence is that one produces the same density of NLSP for lighter masses.
Similar effect can be obtained for the neutralino NLSP. Constraints on dark matter calculated in \eq{eq:OmegaNTP} impose either lower values of $m_{1/2}$ to compensate the increase in $\Omega_{\stau}h^2$ or smaller gravitino masses. For a given mass, the increase in the stau abundance will have important consequences to solve the lithium problems as described in the next section. \fig{Fig:relic_modgrav} represents the evolution of the stau relic density with respect to $\kappa_D$ and $n_D$ for $m_{1/2}=1000 \unit{GeV}$, $m_0=100 \unit{GeV}$ and $\tan \beta=10$. The relic density increases very quickly when the density of the dark component becomes important with $\kappa_D$ and $n_D$. Lines indicate the gravitino relic density compatible with WMAP from NTP only. The dark component contribution must be smaller for increasing gravitino mass.

\subsection{Total gravitino relic density}
\begin{figure}
\includegraphics[width=10cm]{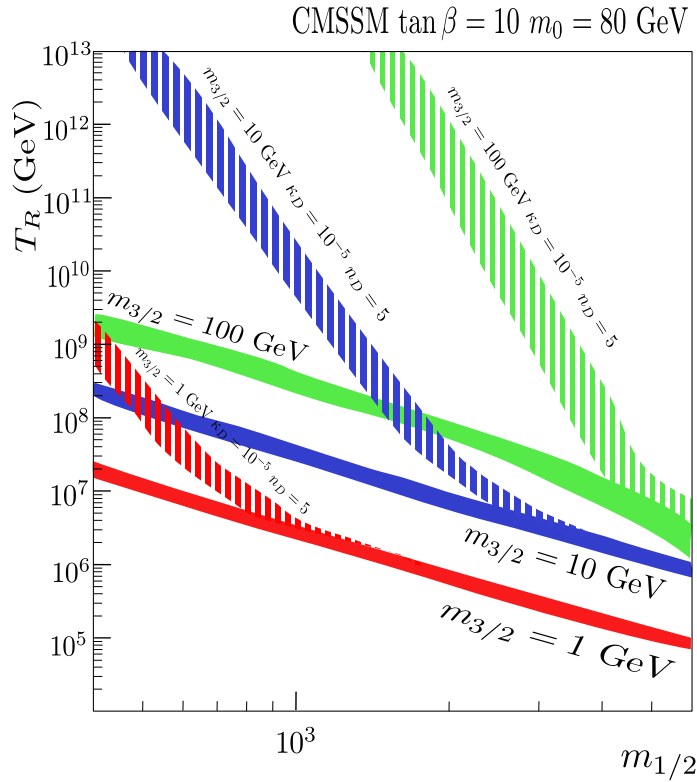}
\centering \caption{\label{fig:reheat}Constraints on reheating temperature for $\Omega_{3/2} h^2$ with both TP and NTP contribution compatible with WMAP measurements, for the standard case and for a dark component with $\kappa_D=10^{-5}$ and $n_D=5$ with gravitino masses $\mgrav=1, 10, 100\unit{GeV}$}
\end{figure}

The total gravitino relic density is the sum of thermal and non-thermal production
\begin{equation}
\Omega_{3/2} h^2 = \Omega_{3/2}^{\rm TP} h^2 + \Omega_{3/2}^{\rm NTP} h^2
\end{equation}
Both contributions are affected by the presence of a dark component acting on the expansion rate in the pre-BBN era. While $\Omega_{3/2}^{\rm NTP}$ increases $\Omega_{3/2}^{\rm TP}$ reduce. \fig{fig:reheat} shows the limit on the reheating temperature in the standard scenario and in the modified expansion rate with $\kappa_D=10^{-5}$ and $n_D=5$. The areas correspond to $\Omega_{3/2} h^2$ compatible with WMAP constraints. For small values of $m_{1/2}$, the relic density is dominated by the thermal production. Higher gravitino masses allow higher values of reheating temperature. In the standard scenario with $m_{3/2}=100\unit{GeV}$, one can observe the beginning of a change in the slope at $m_{1/2}\sim 5 \unit{TeV}$ when NTP becomes dominant and alone reproduces all dark matter putting severe limits on the reheating temperature to ensure negligible thermal production. In the non-standard scenario, bounds on $T_{R}$ are much higher because of the suppression illustrated in \fig{fig:thermal}. The suppression increases with $T_R$ and as the limit on the reheating temperature are higher for low $m_{1/2}$ in the standard case, the bounds in the non-standard model are much higher at low values of $m_{1/2}$. Suppression is less efficient at low reheating temperature leading to a convergence of both scenarios at large values of $m_{1/2}$. Bounds on $T_R$ in the non-standard cosmologal picture are easily in agreement with thermal leptogenesis requirements.

\section{\label{sec:BBN}Lithium problems}

Big Bang Nucleosynthesis describes the production of light elements during the early Universe. The success of the predictions of the standard model of BBN (SBBN) compared to the observations was a strong argument in favour of the hot Big Bang scenario as it implies that the Universe has expanded and cooled down from a very hot state and also suggests that only a small component of matter is baryonic, the rest being mostly dark matter. The SBBN model
depends on a unique parameter, the baryon-to-photon ratio measured by WMAP \cite{Komatsu:2008hk} 
\begin{equation}
\eta_{10}=10^{10}\frac{n_b}{n_\gamma}=6.225 \pm 0.170
\end{equation}
and therefore makes simple prediction on the abundance of light elements. Comparison of predictions and observations are nicely compatible for deuterium and helium-4, and somewhat inconclusive for helium-3 (due to important uncertainties in the post-BBN evolution of this element). However important discrepancies arise for lithium-7 and also for lithium-6. The lithium-7 abundance is inferred from absorption lines in the atmosphere of low-metallicity halo or globular cluster stars. A spectacular result namely the Spite plateau \cite{Spite:1982dd} put forward that, for low-metallicity stars, the $\Li7$ abundance depended linearly with a small slope on metallicity and with very little scatter indicating that depletion would not have been very effective. One can interpret the extrapolation at zero-metallicity as the primordial abundance of $\Li7$. Many observations \cite{Thorburn:1994ib, *Ryan:1999jq, *Charbonnel:2005am, *Hosford:2008rs, Asplund:2005yt} give an abundance in a range of $\Li7/\Hh=(1.1-1.5)\times 10^{-10}$. Some uncertainties on the estimation of the atmospheric temperature of the stars could lead to higher estimations $\Li7/\Hh=(2.19\pm0.28)\times10^{-10}$ \cite{Bonifacio:1996jv, *Bonifacio:2002yx} or even higher values $\Li7/\Hh=(2.34\pm0.32)\times10^{-10}$ \cite{Melendez:2004ni}. The SBBN model prediction is a factor 3-4 higher than the observations. The latest estimations are $\Li7/\Hh = 5.24_{-0.67}^{+0.71}\times 10^{-10}$ \cite{Cyburt:2008kw} and $(5.14\pm0.50)\times10^{-10}$ \cite{Coc:2010zz}. The measurement of lithium-6 has also lead to a possible $\Li6$ problem. SBBN prediction are of order $\Li6/\Hh \sim 10^{-15}-10^{-14}$ \cite{Nollett:1996ef} but substantial amount of $\Li6$ has been measured in stars $\Li6/\Li7\sim 0.05$ \cite{Asplund:2005yt}, most measurements corresponding to $\Li6/\Hh = (3-5)\times 10^{-12}$ \cite{Smith:1993zz, *Hobbs:1997aa}. This lithium-6 issue is still controversial and more data is required to assure that there is a substantial production of lithium-6 during BBN. Here we will assume the possible existence of large amount of lithium-6 conflicting with its standard abundance which should be very low because the only $\Li6$ production reaction in SBBN $\Hd(\alpha,\gamma)\Li6$ is a quadrupole transition. Most $\Li6$ is produced by cosmic ray nucleosynthesis from spallation or cosmic ray fusion, but the abundance of $\Li6$ in low-metallicity stars is too high by two orders of magnitude to be explained by standard cosmic ray nucleosynthesis. The lithium discrepancies could find their origin in numerous sources (for a review see \cite{Lambert:2004kn}). First for lithium-7, the nuclear network used for BBN calculations could have some uncertainties thus overestimating $\Li7$ and underestimating $\Li6$ abundances \cite{Cyburt:2003ae, Coc:2003ce}. It seems very unlikely that the uncertainties could bridge completely the discrepancies. A second hypothesis comes from possible systematic errors in the determination of $\Li7$ and $\Li6$ abundances. Another possibility is that the measured abundance is not primordial but has been reduced by stellar depletion, atmospheric $\Li7$ could be destroyed while transported to inner parts of the stars by nuclear burning. However both the presence of the Spite plateau with little scatter and the presence of large amount of $\Li6$ which is more fragile than $\Li7$ and should be even more depleted argue against a simple depletion mechanism. Some models with turbulent mixing \cite{Vauclair:1998it, *Richard:2004pj} can account for $\Li7$. Combining atomic diffusion and turbulent mixing \cite{Korn:2006tv,*Melendez:2010kw} can deplete efficiently $\Li7$ bringing SBBN prediction and observations in satisfying agreement but is somewhat artificial and ad hoc.

A solution for both lithium problems can be found by modifying SBBN. Relic particles could decay during BBN producing Standard Model particles \cite{Dimopoulos:1987fz, Jedamzik:1999di, Jedamzik:2004er, Kawasaki:2004qu}. These particles would induce non-thermal reactions that would not disturb much the abundance of the other elements but could produce $\Li6$ through reactions such as $\Ht(\alpha,n)\Li6$ or $\He3(\alpha,p)\Li6$ with $\Ht$ and $\He3$ produced from $\He4$ spallation. Non-thermal reactions, solutions for the $\Li7$ problem were also investigated. Early attempts of reduction of $\Li7$ by photo-disintegration \cite{Feng:2003uy} were incompatible with lower limit on $\Hd/\Hh$ or upper bound on $\He3/\Hd$ \cite{Kawasaki:2000qr, Ellis:2005ii}, but injection of neutrons during BBN reduces the $\Li7$ abundance \cite{Jedamzik:2004er}. This scenario has been studied intensively as it was a possible solution to the lithium problems \cite{Jedamzik:2004er, Ellis:2003dn, Feng:2004zu, Feng:2004mt, Roszkowski:2004jd, Cerdeno:2005eu,Jedamzik:2005dh, Pospelov:2006sc, Steffen:2006hw, Pradler:2007is, Bailly:2008yy, Bailly:2009pe}.

In \cite{Bailly:2008yy}, Fig. 2 and Fig. 3 give model independent requirements for the NLSP abundance, hadronic branching ratio and lifetime to solve lithium problems. For the case of a stau NLSP with a typical hadronic branching ratio $B_{\rm had}\sim 10^{-3}$, the abundance must be of order $\Omega_{\tilde{\tau}}h^2\sim 0.5$ corresponding to a stau masses of $\mathcal{O}(1\unit{TeV})$. To get lifetimes of $\tau\sim 3000\unit{sec}$, the gravitino mass must be of order $\mathcal{O}(100\unit{GeV})$. These rough estimates were confirmed in precise calculations in CMSSM or GMSB models \cite{Bailly:2008yy} but this requires a quite heavy SUSY spectrum rendering almost impossible the production of SUSY particles at the LHC.

By using the non-standard cosmological scenario with the addition of the dark component, it is possible to satisfy the previous requirements to solve the lithium problems with lighter stau. From the construction of the dark component model, we impose that radiation dominates during the BBN era. Therefore we do not expect strong variations of the abundance of light elements produced during BBN due to a modified expansion parameter. It is confirmed by Arbey \& Mahmoudi \cite{Arbey:2009gt} in Fig.1 in which the production of $\He4$ and $\Hd$ are close to SBBN values in the correct parameter range for $\kappa_D$ and $n_D$. The authors observe strong productions of these elements for small values of $n_D$ and high values of $\kappa_D$ meaning that the dark component dominates during BBN which conflicts with the requirement of radiation domination during this period.

To solve the lithium problems, we combine the known effects of a long-lived relic particle decaying during Big Bang Nucleosynthesis and the non-standard cosmological scenario with a modified expansion rate. The larger Hubble parameter increases the abundance of relic particles, in our case, the stau NLSP decaying to the gravitino. And because of the larger abundance of stau, it is possible to solve the lithium problems with a lighter stau NLSP and a lighter gravitino.

\subsection{Calculation procedure}

The computation of the production of light elements during BBN must include decaying relic particles and bound state formation between the charged stau and the light elements. These bound states have catalytic effects on BBN and were first stressed by Pospelov \cite{Pospelov:2006sc} focusing on lithium-6 production becoming very efficient through the reaction $(\He4-\stau)+\Hd \rightarrow \Li6 + \tau$. Many other processes were studied \cite{Kaplinghat:2006qr, *Kohri:2006cn, *Cyburt:2006uv, Kawasaki:2007xb, Jedamzik:2007qk, Hamaguchi:2007mp, Bird:2007ge, *Jittoh:2007fr, *Jedamzik:2007cp, *Pospelov:2007js, *Pospelov:2008ta} and full cross-sections were calculated by Kamimura et al. \cite{Kamimura:2008fx}. The BBN computation is based on the SBBN Kawano code and modified by K. Jedamzik \cite{Jedamzik:2006xz, Jedamzik:2007qk} to include all the latest improvements on decaying relic particles and bound states. The Jedamzik code requires a number of parameters: the hadronic and electromagnetic branching ratios for the decay of the NLSP and the energies in the cascades. The NLSP decays to the gravitino and various Standard Model particles such as photons, electrons and positrons (electromagnetic cascades) or neutron and protons (hadronic cascades). These particles interact with the light elements present in the plasma inducing reactions like photo-dissociation, spallation\ldots

The stau decay is mostly dominated by the two-body decay $\stau\rightarrow \tau \tilde{G}$ therefore its lifetime is given by the decay width of this process $\tau=\hbar/\Gamma(\stau\rightarrow \tau \tilde{G})$
\begin{equation}
\Gamma(\tilde{\tau}\rightarrow \tau \tilde{G}) = \frac{1}{48\pi}\frac{m_{\tilde{\tau}}^5}{M_{\rm Pl}^2 m_{3/ 2}^2} \left( 1-\frac{m_{3/2}^2}{m_{\tilde{\tau}}^2}\right)^4
\end{equation}

Electromagnetic cascades dominate the whole decay process induced mostly by the unstable lepton tau produced in two-body decay. Below $\tau\sim 100 \unit{s}$, the mesons produced in the decay of the lepton tau induce  charge-exchange reactions increasing the neutron-to-proton ratio and the helium-4 abundance (\cite{Kawasaki:2004qu} and references therein). Such effects are not taken into account and therefore, only lifetime above $\tau\sim 100 \unit{s}$ are considered. Above a NLSP lifetime of 100 s, mesons mostly decay electromagnetically.

Hadronic cascades require the production of a quark-antiquark pair that will then hadronize and produce protons and neutrons (for which the energy spectrum is computed using PYTHIA \cite{Sjostrand:2006za}). Therefore 4-body diagrams are required for stau hadronic decay width calculation. The hadronic decay is therefore sub-leading with respect to the electromagnetic decay.

The electromagnetic branching ratio and the energy (simply given from kinematics) are
\begin{equation}
B_{\rm em} = 1-B_{\rm had}\simeq 1 \qquad \mbox{and} \qquad E_{\rm em}=\alpha \pfrac{m_{\rm NLSP}^2 - \mgrav^2}{2m_{\rm NLSP}}
\end{equation}
The lepton tau produced in the decay of its superpartner is unstable and decays producing lighter leptons and neutrinos. The part of energy taken by the neutrinos has no or little effect on the plasma, the coefficient $\alpha$ signs this loss of energy and following \cite{Feng:2003uy} we have $1/3\leq \alpha \leq 1$. We take $\alpha=1/2$.

\begin{figure}
\begin{center}
\includegraphics[width=7cm]{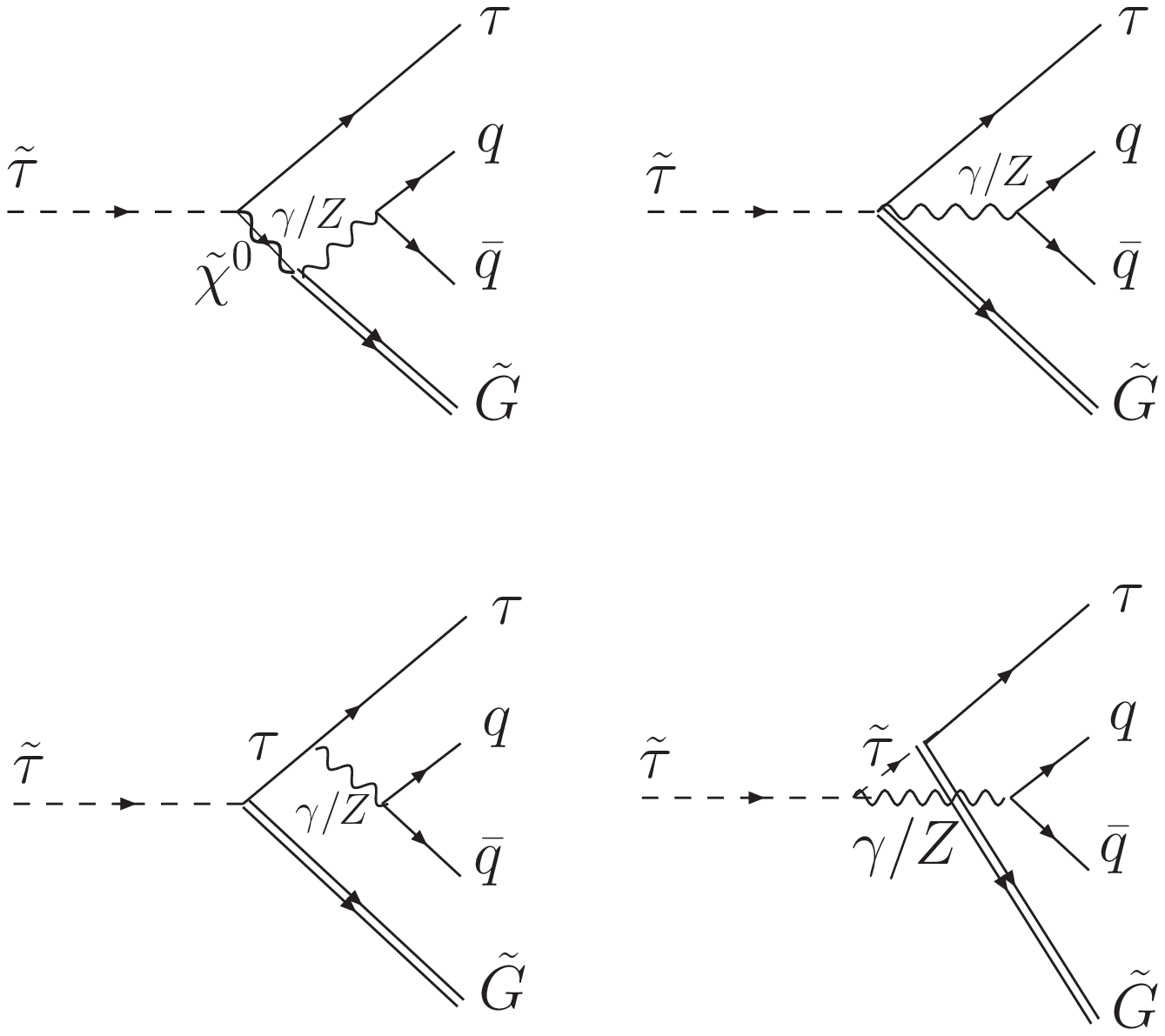}
\qquad
\includegraphics[width=7cm]{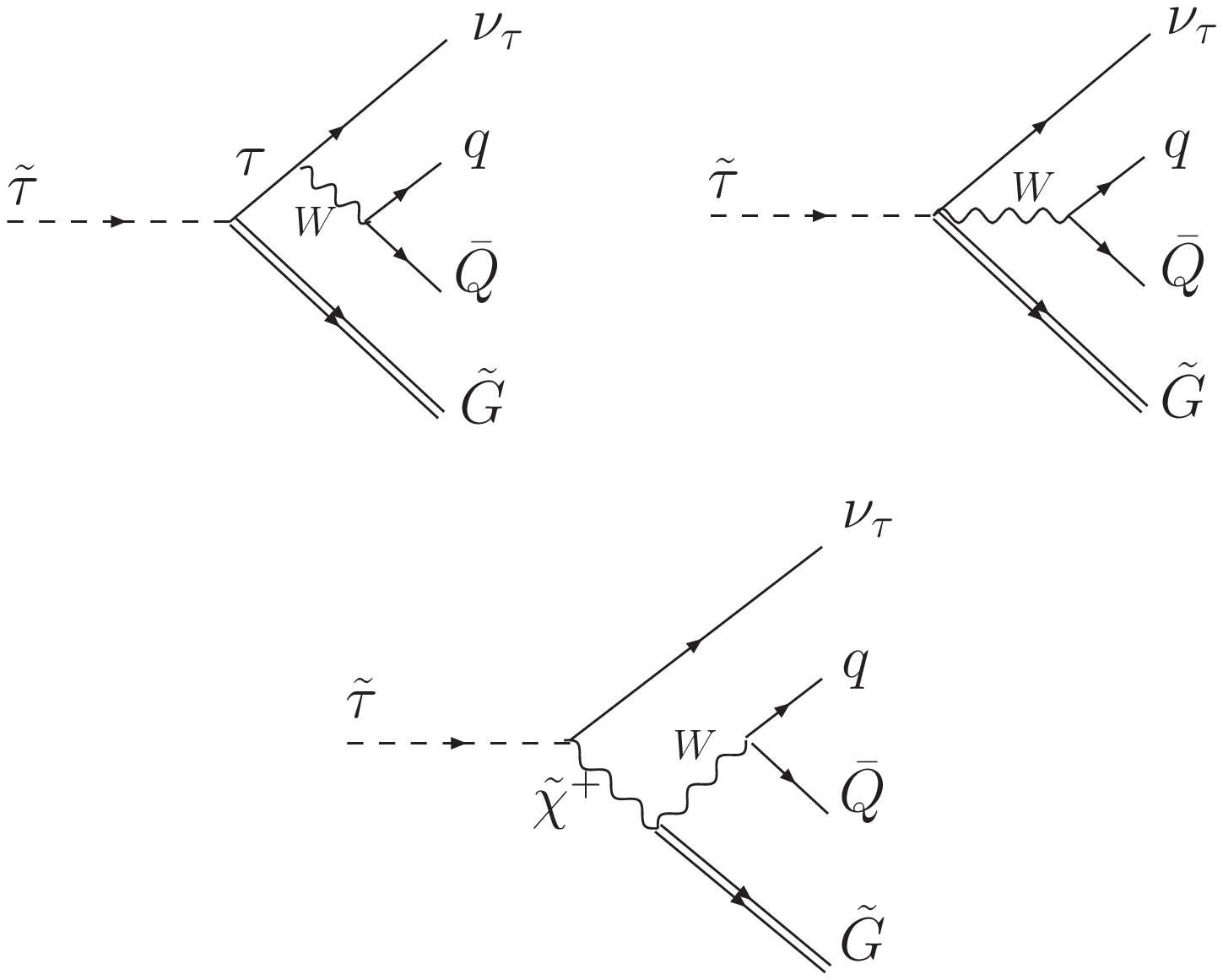}
\end{center}
\caption{\label{Fig:diagrams}Diagrams calculated in the 4-body decay of stau to gravitino}
\end{figure}

The hadronic branching ratio is calculated following \cite{Steffen:2006hw}
\begin{equation}
B_{\rm had}(\stau\rightarrow \tau \tilde{G} q\bar{q};m_{q\bar{q}}^{\rm cut})=\frac{\Gamma(\stau\rightarrow \tau\tilde{G}q\bar{q})}{\Gamma_{\rm tot}}
\end{equation}
with the 4-body decay width
\begin{equation}
\Gamma(\stau\rightarrow \tau\tilde{G}q\bar{q};m_{q\bar{q}})=\int_{m_{q\bar{q}}}^{\mstau-\mgrav-m_\tau} dm_{q\bar{q}}\frac{d\Gamma(\stau\rightarrow \tau \tilde{G} q\bar{q})}{dm_{q\bar{q}}}
\end{equation}
where $m_{q\bar{q}}=2\unit{GeV}$ is a cut on the invariant mass of the pair quark-antiquark corresponding to twice the mass of a nucleon below which hadronization does not produce any nucleon. The decay width is calculated using CalcHEP \cite{Pukhov:2004ca} as in \cite{Bailly:2008yy}. Diagrams taken into account are given in \fig{Fig:diagrams}. The $Z$ boson  width is taken as a Breit-Wigner function. Diagrams with $\stau_2$, neutralinos $\chi_{2}^0$, $\chi_3^0$, $\chi_4^0$ and charginos $\chi_2^\pm$ are neglected assuming these particles are very massive. Diagrams with the exchange of a virtual neutralino or chargino are negligible for $m_{\chi} \gtrsim 1.1 \mstau$ but are always included in the calculation. Contributions to the hadronic branching ratio from diagrams with a virtual $W$ are one to two orders of magnitude smaller than the contribution of processes with a virtual $Z$ because the stau is dominated by its right component. Processes with Higgses are neglected because of the masses in the propagator and the small couplings. In our calculations, we fully take into account the left-right mixing of $\stau_1$ and gaugino mixing of $\chi_1^0$.

The hadronic energy calculation is done following again \cite{Steffen:2006hw} using a mean value based on a convolution of the branching ratio with the quark-antiquark invariant mass
\begin{equation}
E_{\rm had}=\frac{1}{\Gamma(\stau\rightarrow \tau\tilde{G}q\bar{q})}\int_{m_{q\bar{q}}}^{\mstau-\mgrav-m_\tau} dm_{q\bar{q}} \; m_{q\bar{q}}\frac{d\Gamma(\stau\rightarrow \tau \tilde{G} q\bar{q})}{dm_{q\bar{q}}}
\label{eq:Ehad}
\end{equation}
It was mentioned in \cite{Bailly:2008yy} that this estimation of the hadronic energy had some problems. First the energy is estimated in the rest frame of the quark-antiquark pair and not in the rest frame of the stau. This calculation does not allow to estimate the boost needed to change from one frame to another. Also the fact that this is a mean energy disregards completely the fact that one nucleon does not have the same impact on the plasma as ten nucleons with a total energy equal to the one nucleon. A more realistic generation of nucleons and their energy in the rest-frame of the stau was used with CalcHEP and PYTHIA yielding an estimation of the uncertainties on the abundance of light elements. This study resulted in 10-20\% uncertainty for deuterium, 30-40\% for lithium-7 and 20\% for lithium-6. In this paper we will use \eq{eq:Ehad} keeping these uncertainties in mind.

\subsection{BBN results}

BBN calculation can now be performed taking into account a great number of effects: SBBN reactions, non-thermal reactions from the stau decay, bound state formation and the modified expansion rate. Abundance of deuterium, helium-3, helium-4, lithium-6 and lithium-7 are estimated and confronted to observational constraints. These constraints are taken on conservative basis to include observational uncertainties, post-BBN evolution and theoretical uncertainties in some reaction rates and hadronic energy calculation. We take the following BBN constraints :
\begin{eqnarray}
1.2\times 10^{-5} \leq & \Hd/\Hh & \leq 5.3\times 10^{-5}\\
& \He3/\Hd & \leq 1.72 \\
& Y_p & \leq 0.258 \\
8.5\times 10^{-11} \leq & \Li7/\Hh & \leq 2.5\times 10^{-10}\\
0.015 \leq & \Li6/\Li7 & \leq 0.66
\end{eqnarray}
The last two constraints are the abundances satisfying lithium-7 and lithium-6 abundances inferred by observations. In the following figures, SBBN values for lithium isotopes will be taken as
\begin{eqnarray}
2.5\times 10^{-10} \leq & \Li7/\Hh & \\
& \Li6/\Li7 & \leq 0.015
\end{eqnarray}

\begin{figure}
\centering \includegraphics[width=10cm]{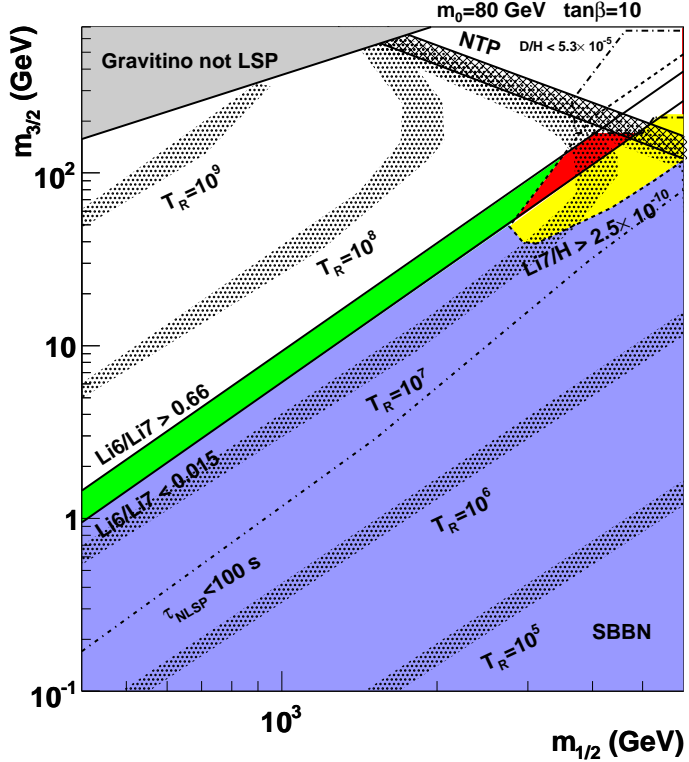}
\caption{\label{fig:BBNstandard}Big Bang Nucleosynthesis results in CMSSM with $m_0=80\unit{GeV}$ and $\tan\beta=10$. The light blue region produces light elements abundances compatible with SBBN, masses solving the lithium-7 discrepancy (yellow), lithium-6 (green) and both problems (red). The hatched area labelled NTP indicates a gravitino relic density from non-thermal production alone compatible with measurements from WMAP and the other hatched correspond to the sum of TP+NTP for given reheating temperature. A dashed line also indicates the 100 s NLSP lifetime limit.}
\end{figure}

\fig{fig:BBNstandard} presents the light element abundances in the standard cosmological scenario with the CMSSM model and $m_0=80\unit{GeV}$ and $\tan\beta=10$. Gaugino mass is chosen such that the stau is the NLSP so $m_{1/2}=[400-6000]\unit{GeV}$ and the gravitino mass $m_{3/2}=[10^{-1}-700]\unit{GeV}$. Strictly speaking, in the CMSSM the gravitino mass is of the order of the electroweak scale and lighter masses are more typical of gauge mediated supersymmetry breaking scenarios. It was shown in \cite{Bailly:2008yy} that the choice of mediation mechanism did not change much the phenomenology of BBN therefore we use the CMSSM model and assume that the gravitino mass is a free parameter. The limits from deuterium, $\Li6$ and $\Li7$ \fig{fig:BBNstandard} delineate different regions. Light blue corresponds to values of $m_{1/2}$ and $m_{3/2}$ satisfying the SBBN values. Yellow indicates masses solving the $\Li7$ problem, green region satisfies the $\Li6$ corresponding to stau lifetime of $1600-6300\unit{sec}$ for an upper limit of $\Li6/\Li7<0.66$. For a more stringent limit of $\Li6/\Li7<0.15$, the stau lifetime does not exceed $4\times 10^3 \unit{sec}$. Above the $\Li6/\Li7>0.66$ limit, lithium-6 is overly produced because of bound state catalytic effects mainly the $(\He4-\stau)+D\rightarrow \Li6+\stau$ reaction. The red region solves both problems simultaneously for $m_{1/2}=[3-5]\unit{TeV}$ corresponding to stau masses $\mstau=[1-1.8]\unit{TeV}$ and gravitino masses $\mgrav={60-120}\unit{GeV}$. Note that for such heavy spectrum, the gluino has a mass of almost 10 TeV which makes impossible the production of such particle in the LHC. For higher values of $\tan \beta$ would reduce the relic density of stau and therefore would require much more massive sleptons to produce the same impact of BBN.

The hatched zone labelled NTP corresponds to the gravitino relic density coming only from non-thermal production compatible with WMAP constraints for dark matter. If one adds the contribution from thermal production, different regions satisfy the WMAP values depending on the reheating temperature. For $T_R \lesssim 10^{7} \unit{GeV}$, it is possible to satisfy to SBBN and dark matter constraints. In regions solving lithium problems, all dark matter is reproduced by gravitino relic density for $T_R \sim  10^{7} \unit{GeV}$. For smaller values of the reheating temperature, gravitino relic density lies below dark matter relic density and one must assume a mixed dark matter scenario with the gravitino and another candidate. At reheating temperatures $T_R \gtrsim 10^{8}\unit{GeV}$, constraints on gaugino mass and gravitino mass which must be heavier to limit the effect of thermal production do not permit to solve BBN constraints and are therefore excluded.

\begin{figure}
\centering \includegraphics[width=7.5cm]{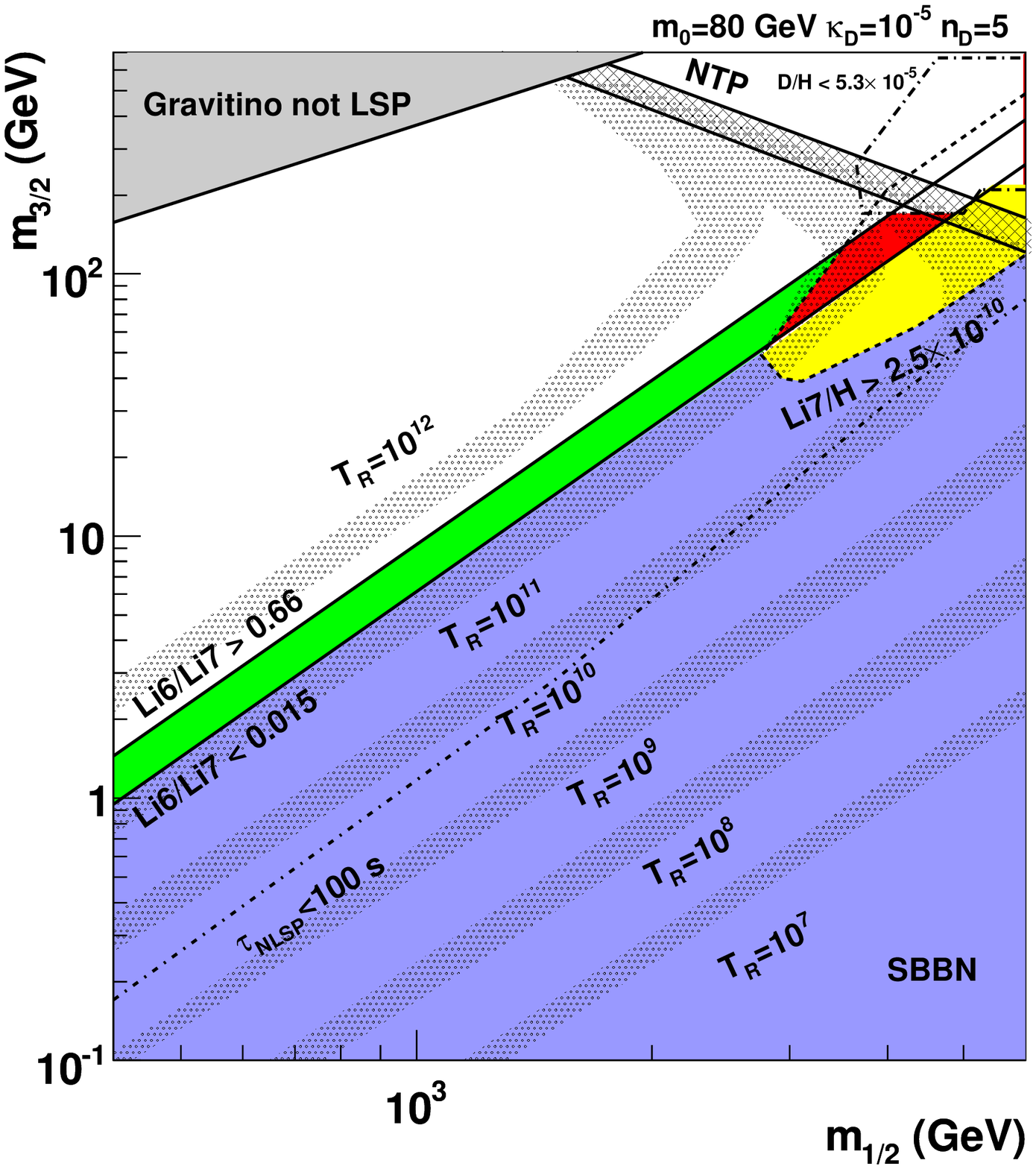}
\centering \includegraphics[width=7.5cm]{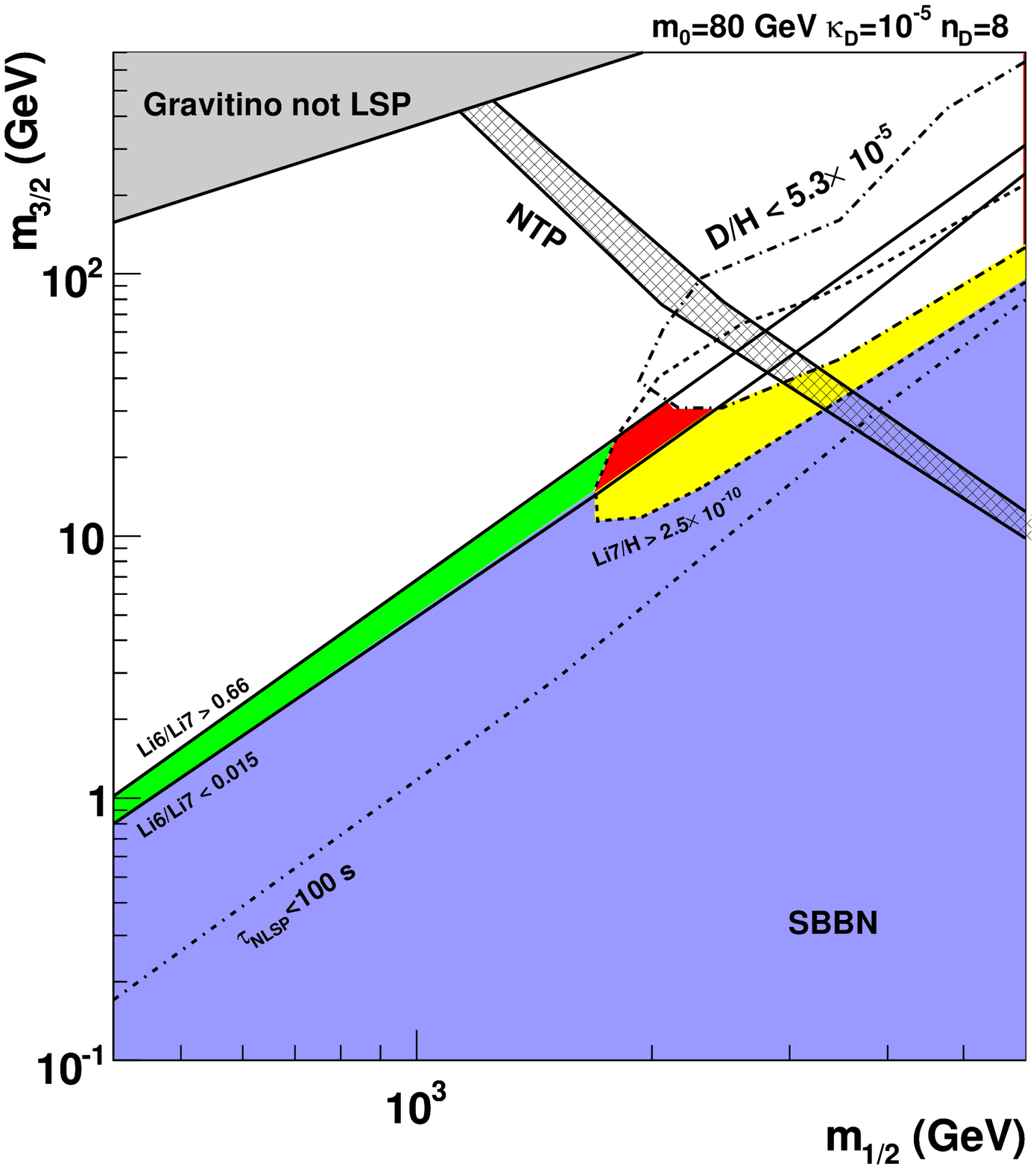}
\caption{\label{fig:BBNnonstandard}Results for Big Bang Nucleosynthesis in the modified expansion rate scenario with $\kappa_D=10^{-5}$ and $n_D=5$ (left) or $n_D=8$ (right). Colours and hatched areas defined as in \fig{fig:BBNstandard}.}
\end{figure}

Similar results are presented \fig{fig:BBNnonstandard} in a non-standard scenarios with the dark component. We use the same CMSSM parameters and the dark component are taken to be $\kappa_D=10^{-5}$ and $n_D=5$ (left) and $n_D=8$ (right). For $n_D=5$, the effect of the dark component is strong enough to suppress the thermal production and relax the reheating temperature upper bound as mentioned before. It is now possible to solve both the lithium and dark matter problems with a reheating temperature up to $T_R\sim 10^{11}\unit{GeV}$, 4 orders of magnitude above the limit in the standard case, as indicated \fig{fig:reheat} (for a gravitino mass $m_{3/2}=1\unit{GeV}$). For this scenario, effects on BBN is similar to what was found in the standard scenario. The dark component induces an early freeze-out of the stau and therefore increases its abundance but for these parameters, $\kappa_D=10^{-5}$ and $n_D=5$, the variation on the stau abundance differs of less than a percent and the BBN results are unchanged.

The figure on the right has an index more important $n_D=8$ leading to an energy density of dark component much more important. The consequence on the thermal production is a very strong suppression illustrated \fig{fig:reheat} (5 orders of magnitude at $T_R=10^5\unit{GeV}$ and the suppression increases with $T_R$). Only very little abundance of gravitino is formed and can be neglected with respect to the non-thermal production. There are no constraints on the reheating temperature any more. And one can choose any reheating temperature to be compatible with for instance thermal leptogenesis.

On the other hand stau production is very efficient, enhanced by two orders of magnitude (see \fig{Fig:relic_modgrav}), because of the early freeze-out. The gravitino relic density from the non-thermal production is increased by the same factor. For BBN, the production of light elements changes strongly because much more unstable relic particles are present in the plasma. Upper bound on deuterium and lower bound on lithium-7 are reached for smaller values of $m_{1/2}$. Bound state effects are also much more important since the abundance of stau is higher and limits on lifetime are much more severe: $\tau<2500\unit{sec}$ for $\Li6/\Li7<0.66$ or $\tau < 2000\unit{sec}$ for $\Li6/\Li7<0.15$. Abundances compatible with observations and the resolution of the lithium problems can be obtained with lighter stau than in the standard scenario $\mstau \sim 600-700 \unit{GeV}$ and gravitino masses $\mgrav\sim 20\unit{GeV}$. In this specific case of parameter choice for the dark component, it is not possible to solve both lithium problems and reproduce dark matter relic from gravitino alone. It is however possible to solve the lithium-7 problem and satisfy dark matter. In the region solving both lithium problems, the thermal production is negligible and the non-thermal production contributes to 10-20\% of the dark matter relic density. One must keep in mind that dark matter could be mixed with another component.

\section{Conclusion}

Supersymmetric models with gravitino LSP and stau NLSP have interesting applications for two cosmological issues, namely dark matter and Big Bang Nucleosynthesis. The gravitino production from scattering processes during reheating and the decay of the NLSP can account for all dark matter in the Universe. And moreover, the long-lived stau decay brings solutions to both lithium problems. This nice set-up requires a very heavy spectrum and a quite low reheating temperature.

In this paper, we have supposed that the expansion rate of the Universe previous to BBN differs from the standard picture because of the presence of a dark component. The consequences are various. On the one hand the thermal production of gravitino can be strongly suppressed relaxing the constraints on the upper bound for the reheating temperature. On the other hand stau freeze-out occurs earlier leading to a higher abundance for the slepton and a more important gravitino non-thermal production. And because of the increase in the stau abundance, equivalent impact on BBN as in the standard case can be reached with a lighter stau.

The very heavy spectra of the standard scenario is completely out of reach of the LHC, but the non-standard cosmology yield a lighter spectrum that may be investigated at the LHC. We have illustrated the case $\kappa_D=10^{-5}$ and $n_D=8$ where $\mstau \sim 600-700 \unit{GeV}$ which could be still quite hard to detect since squark masses are above $2\unit{TeV}$. If $\kappa_D=10^{-8}$, $\mstau\sim 400 \unit{GeV}$, supersymmetric events could be produced at the LHC running at $14\unit{TeV}$. Detection was studied for instance in \cite{Ellis:2006vu, Tarem:2009zz}. Every supersymmetric event would induce a cascade ending with the lightest stau which will be stable on the scale of the detector (lifetime $\tau\sim 10^3 \unit{sec}$) and would be identified as a heavy muon with low velocities and distinctive time-of-flight and energy-loss. The signature is quite clean to be distinguished from background and detection could give a measurement of the stau mass with its consequences for our dark matter and BBN scenario, illustrating a nice complementarity between cosmology and collider physics.

\section*{Acknowledgements}
I would like to thank K. Jedamzik for the possibility of using his BBN code, Genevi\`eve B\'elanger, Gilbert Moultaka, Richard Taillet, R\'emi Lafaye for helpful discussions and Anthony Hillairet for useful comments. This work was supported in part by the ANR project ToolsDMColl, BLAN07-2-194882.

\bibliographystyle{h-physrev2}
\bibliography{biblio}{}

\end{document}